\documentclass{IEEEoj}
\usepackage{cite}
\usepackage{amsmath,amssymb,amsfonts}
\usepackage{algorithmic}
\usepackage{graphicx,color}
\usepackage{textcomp}
\usepackage{hyperref}

\AtBeginDocument{\definecolor{ojcolor}{cmyk}{0.93,0.59,0.15,0.02}}
\begin{document}

\title{Exploring Best Practices for ECG Pre-Processing in Machine Learning}

\author{Amir Salimi\authorrefmark{1}, Sunil Vasu Kalmady\authorrefmark{2}, Abram Hindle.\authorrefmark{1,2},
Osmar Zaiane, Padma Kaul}
\affil{University of Alberta, Computing Science, Edmonton, Alberta, Canada}
\affil{University of Alberta, Canadian Vigour Centre, Edmonton, Alberta, Canada}
\corresp{CORRESPONDING AUTHOR: Amir Salimi (e-mail: asalimi@ualberta.ca).}
\authornote{This work has been funded by the Canadian Institutes of Health Research and the Heart and Stroke Foundation of Canada.}
\markboth{ECG Signal Processing in Machine Learning}{Salimi \textit{et al.}}

\begin{abstract}
In this work we search for best practices in pre-processing of Electrocardiogram (ECG) signals in order to train better classifiers for the diagnosis of heart conditions.
State of the art machine learning algorithms have achieved remarkable results in classification of \textit{some} heart conditions using ECG data, yet there appears to be no consensus on pre-processing best practices. Is this lack of consensus due to different conditions and architectures requiring different processing steps for optimal performance? Is it possible that state of the art deep-learning models have rendered pre-processing unnecessary? In this work we apply down-sampling, normalization, and filtering functions to 3 different multi-label ECG datasets and measure their effects on 3 different high-performing time-series classifiers. We find that sampling rates as low as 50Hz can yield comparable results to the commonly used 500Hz. This is significant as smaller sampling rates will result in smaller datasets and models, which require less time and resources to train. Additionally, despite their common usage, we found min-max normalization to be slightly detrimental overall, and band-passing to make no measurable difference. We found the blind approach to pre-processing of ECGs for multi-label classification to be ineffective, with the exception of sample rate reduction which reliably reduces computational resources, but does not increase accuracy.

\end{abstract}
%


\maketitle

\section{Introduction}
\label{sec:intro}
We focus on the Electrocardiogram (ECG) pre-processing decisions faced by researchers working towards automatic classification of cardiovascular conditions. The ever-growing number of tools and approaches available for machine learning often require a number of a priori guesses to be made by those who process the training data and design the architectures. The task of training models which predict heart disease is no exception.

Routinely utilized by clinicians for diagnoses of cardiovascular abnormalities, an ECG is a recording of the electrical activity of the cardiovascular system~\cite{maleki2014use,uwaechia2021comprehensive}. In recent years, machine learning models have achieved remarkable results in automatic diagnosis of some heart conditions when trained with enough labeled ECG data~\cite{reyna2021will,reyna4issues,ribeiro2020automatic,chen2020detection,zhao2020adaptive}. However, training such models requires a large amount of data~\cite{reyna2021will,reyna4issues,zhao2020adaptive}, where decisions regarding how the data is pre-processed (e.g., filtering, scaling, data augmentation) can be critical for the model's performance, hardware requirements, and training time.

 Pre-processing functions are transformations applied to signals---ECGs, in this case---in order to reduce noise and simplify the learning task. Previous works utilize a diverse range of pre-processing functions, datasets, and architectures, often with great results~\cite{hong2022practical}. As we will discuss in Section~\ref{sec:prevwork}, the variety of approaches in past works makes finding a trend between the pre-processing functions used and performance results not feasible. We believe that in order to make claims about the viability of ECG pre-processing methods (that is, whether they result in better outcomes) it is best to consider multiple datasets, architectures, and heart conditions.  

Here, we aim to simplify this decision space for other researchers by testing the effect of \textit{down-sampling}, \textit{band-passing}, and \textit{normalization} on the outcome of 3 different cutting-edge multi-label classifiers for time-series when trained on 3 different ECG datasets. We show that down-sampling from the standard recording rate of 500Hz can significantly reduce training times and hardware requirements, without being detrimental to model performance.  We found min-max normalization to be slightly detrimental overall, and band-passing to make no measurable difference. In Section~\ref{sec:experiment} we discuss the experiment setup and performance measures, and discuss the results of down-sampling, band-passing, and normalization in Sections~\ref{sec:scaling},~\ref{sec:BandPass}, and~\ref{sec:Norm}. Our methodology and datasets can be replicated using our code, which is publically available\footnote{\url{https://github.com/imilas/ecg_augmentation}}.

\begin{table*}[htbp]

\caption{Functions used by Top 5 teams in Physionet2020~\cite{hong2022practical}}%
\centering
    
    \begin{tabular}{|c|c|c|c|c|c|}
    \hline
    Rank & Group                                          & Scaling                   & Normalize            & BandPass  & Window              \\ \hline
    1    & Wide and Deep Transformer~\cite{natarajan2020wide}                   &                           & \checkmark & \checkmark & 15s, 0 Pad\\ \hline
    2    & Adaptive ResNet~\cite{zhao2020adaptive}                  & \checkmark (257 Hz) & \checkmark & & 16s, 0 Pad\\ \hline
    3    & SE-ResNet~\cite{zhu2020classification}        &                           &                           &                           & 30s, 0 Pad\\ \hline
    4    & Scatter Transform and DNNs ~\cite{oppelt2020combining}             &                           & \checkmark &                           & 20.5s \\ \hline
    5    & Adversarial Domain Generalization~\cite{hasani2020classification} &    \checkmark (250 Hz) & \checkmark   & \checkmark & \\ \hline
    \end{tabular}
    \label{tab:top5}
\end{table*}
\section{Previous Work}
\label{sec:prevwork}
The state-of-the-art benchmarks for the automatic classification of cardio-vascular diseases continually improve as more datasets become available and computation costs decrease. However, no consensus seems to exist on the best pre-processing steps for such datasets. In this work we test various approaches to pre-processing taken by previous works in-order to measure their effects on classification performance. 

Here, we define pre-processing as \textit{``any function which aims to simplify or transform the data in-order to enhance the performance outcomes and/or reduce training over-head"}. \cite{hong2022practical} highlight 4 common pre-processing methods for ECG signals, based on their analysis of entries to the Physionet2020 Challenge~\cite{goldberger2000physiobank,alday2020classification}: resizing (which we do not experiment on), resampling, band-passing, and normalizing. Examples of works using one or more of these methods are given in Table~\ref{tab:top5}. 

\subsection{Resizing} 
Signals are time-series, and time-series vary in length. Resizing refers to assigning a fixed length (or window) to all signals. Here, we use a fixed window length of 5000 (10 seconds) for all experiments, and pad signals with zeros on the right, if the full size $x$ is lower than our desired fixed length. If $x$ is longer than the fixed length, then samples 0-$x$ are used. 

\subsection{Resampling (Down-Sampling):}
ECGs are typically sampled at the rate of 500Hz~\cite{luo2010review,uwaechia2021comprehensive}, meaning that each second of an ECG recording contains 500 samples. Down-sampling an ECG is often no different than resolution reduction of an image. Since a lower sampling rate leads to smaller data, it can often drastically lower training times and hardware requirements, but important information may be lost. In this work we use the simple and commonly used nearest-neighbour interpolation~\cite{rukundo2012nearest} method to measure the effect of different sampling rates on the performance of ML models. We discuss our findings with regards to performance and model sizes in Section~\ref{sec:scaling}. 

\subsection{Band-Passing}
Many sources can introduce noise to ECG signals. Powerlines are a common example~\cite{uwaechia2021comprehensive}, which depending on the regional infrastructure can introduce noise at various frequencies. It is common for ECG recording hardware to use built-in band rejection (or notch filters) to deal with such issues, and many ECG libraries apply notch filters to ECGs for this reason~\cite{Makowski2021neurokit}. However, there are other unaccounted sources of noise which need to be dealt with at the software level. In signal processing, a band-pass filter is used for the removal of frequencies outside of a given range (or cut-offs)~\cite{lyons1997understanding}; all frequencies below the high-pass cut-off and above the low-pass cut-off are removed or reduced. This can improve performance as noise can interfere with the learning process. In this work we compare the use of different cut-offs as well as the case when no filtering was applied to the dataset. The results are shown in Section~\ref{sec:BandPass}. 

\subsection{Normalizing}
Normalization of input values has been shown to produce better or equal value results in image and time-series classification~\cite{bhanja2018impact}. Particularly in the case of ECGs, there can be differences in recorded amplitudes depending on the equipment used~\cite{uwaechia2021comprehensive}, making normalization a logical step in pre-processing. Here we compare the performance of min-max normalization to raw ECGs and show the results in Section~\ref{sec:Norm}.
\section{Models}
There is a large variety in the architectures used to learn from ECG datasets, or time-series data more generally. Here, we use three models with proven results in state of the art time-series classification benchmarks, implemented by the TSAI library~\cite{tsai}. The \textbf{Inception-Time Network}, which is the 1-dimensional application of the Inception-Network~\cite{szegedy2017inception,ismail2020inceptiontime} and \textbf{MiniRocket}, a quick and mostly deterministic feature extractor for time-series~\cite{dempster2021minirocket}. A recent survey of time-series classification methods by~\cite{ruiz2021great} highlights both of these models as excellent performers in various multi-variable time-series classification benchmarks. In addition, Inception-Time has excelled in ECG classification benchmarks~\cite{Strodthoff2021}. The third architecture tested here is \textbf{xresnet1d101}, a resnet model for time-series which was found to be the overall best performer in a recent benchmark study of best architectures for classification of ECGs by~\cite{strodthoff2020deep}.
\section{Datasets}
\label{datasets}

We conduct our experiments on three datasets, the~\textbf{CPSC} dataset~\cite{liu2018open}, \textbf{Chapman-Shaoxing} dataset~\cite{zheng202012}, and ~\textbf{PTB-XL}~\cite{wagner2020ptb}. The datasets are all multi-label, and recorded at the sampling rate of 500Hz. All have been released as part of the 8 datasets of labeled 12-lead ECGs provided by the Physionet\-2021 challenge~\cite{reyna2021will,reyna4issues}.
To simplify our datasets and experiments, we only use the labels which appear in more than 5\% of the ECGs in each dataset. Here we only use 10 seconds of data from each ECG. ECGs shorter than this length are right padded with zeros, therefore we can represent each ECG as a matrix with dimensions of 12x5000 (before down-sampling).
The breakdown of these datasets and their label counts after our modifications are given in Tables~\ref{tab:cpsc},~\ref{tab:chapman}, and~\ref{tab:ptb}.



\begin{table}[htbp]
\centering

  \caption{Modified CPSC Dataset}%

    \begin{tabular}{|c|r|}
     \hline
    Label & Count \\
     \hline
    right bundle branch block    &  1857 \\
    ventricular ectopics         &   700 \\
    atrial fibrillation          &  1221 \\
    1st degree av block          &   722 \\
    premature atrial contraction &   616 \\
    sinus rhythm                 &   918 \\
    st depression                &   869 \\
     \hline
    \textbf{Total} & \textbf{6903}\\
    \hline
    \end{tabular}
    \label{tab:cpsc}%
\end{table}

\begin{table}[htbp]
\centering
  \caption{Modified Chapman Dataset}
    \begin{tabular}{|c|r|}
     \hline
    Label & Count \\
     \hline
    left ventricular high voltage &  1295 \\
    atrial fibrillation           &  1780 \\
    t wave abnormal               &  1876 \\
    sinus bradycardia             &  3889 \\
    supraventricular tachycardia  &   587 \\
    sinus rhythm                  &  1826 \\
    sinus tachycardia             &  1568 \\
    nonspecific st t abnormality  &  1158 \\
     \hline
    \textbf{Total} & \textbf{9910}\\
    \hline
    \end{tabular}
    \label{tab:chapman}%
\end{table}

\begin{table}[htbp]
\centering
 
  \caption{Modified PTB-XL}
  
    \begin{tabular}{|c|r|}
     \hline
    Label & Count \\
     \hline
        left axis deviation                  &   5146 \\
        myocardial ischemia                  &   2175 \\
        myocardial infarction                &   5261 \\
        left ventricular hypertrophy         &   2359 \\
        ventricular ectopics                 &   1154 \\
        atrial fibrillation                  &   1514 \\
        t wave abnormal                      &   2345 \\
        abnormal QRS                         &   3389 \\
        sinus rhythm                         &  18092 \\
        left anterior fascicular block       &   1626 \\
        incomplete rbbb &   1118 \\
     \hline
    \textbf{Total} & \textbf{21311}\\
    \hline
    \end{tabular}
     \label{tab:ptb}%
\end{table}

\subsection{CPSC} Released in 2018 as part a multi-label ECG classification competition~\cite{liu2018open} and used in recent benchmarks~\cite{strodthoff2020deep}. ECG signal duration in this dataset is between 6 and 60 seconds, with an average duration of 15.79 seconds. Originally, this dataset has 9 labels, reduced to 7 for our work.

\subsection{Chapman-Shaoxing} Has not been subject to many benchmarks. This dataset contains 10,646 12-lead ECGs, originally with 11 labels, reduced to 8 in our work~\cite{zheng202012}.

\subsection{PTB-XL} Used in recent benchmarks~\cite{strodthoff2020deep}, this dataset contains 21,837 12-lead ECGs~\cite{wagner2020ptb}. The original dataset has 54 labels, reduced to 11 in our work.

\section{Experiment Methodology}
\label{sec:experiment}

In Section~\ref{sec:prevwork} we discussed common pre-processing functions, and how some high-performing works have used one or more of these functions, while others have omitted them entirely (as seen in Table~\ref{tab:top5}). We discussed the three high-performing time-series architectures which we will be using, and in Section~\ref{datasets}, we described the three datasets of multi-labeled ECGs that the models can learn from. Best practices for ECG pre-processing could be established if we observe consistent patterns of improvement in performance scores when applying a pre-processing function to different datasets, and comparing the performances of different models.

To compare pre-processing types, we apply the methods we want to analyze to each dataset, and train models on the transformed datasets. We use the performance of the model to score and rank the pre-processing function. Given enough of these performance scores for the various models, datasets, and diseases, we use statistical tests such as Wilcoxon signed rank test and Kruskal-Wallis to determine whether the pre-processing method was a significant factor in the performance of the models. For example, we would have the CPSC dataset at various sampling rates: CPSC\_500Hz, CPSC\_250Hz, etc. For each version of the dataset (i.e, the dataset pre-processed using a different function), we train and test each model 20 times. For 20 rounds, the data is randomly split in 80/10/10 training/validation/testing sets. The same seeding is used for the random splits such that we can use pair-wise tests to compare experiment results. Each model is trained on the training set, at the end of each epoch the new weights are saved to disk if the model has improved its highest F1 score on the validation set. When no improvements in the F1 measure (on the validation set) is seen after 30 epochs, the training stops, and the highest scoring weights (based on validation set performance) are loaded in the model and the F1 score on the test set is saved as the performance score of the experiment. This means that for each dataset/pre-processing/model combination, we have 20 performance scores. 

In this case, we can use the Wilcoxon signed rank test to measure whether each model out-performs every other model. To rank and compare the models, we use the Wilcox-Holm post-hoc analysis used for creating critical-difference diagrams by~\cite{IsmailFawaz2018deep}. This analysis uses a pairwise Wilcoxon test with Holm-Bonferroni p-value correction, and shows whether the differences between each model pair is significant. While~\cite{IsmailFawaz2018deep} use the average rank to show relative performance, we use the ``Mean Reciprocal Rank" (MRR) for each model. Here, $rank_i$ refers to the model's rank relative to the other models when trained/validated/tested on the same subsets of the data, and $Q=20$ since for each model/dataset combination we have 20 F1 scores:
\[ \text{MRR} = \frac{1}{Q}\sum_{i=1}^{Q}\frac{1}{rank_i} \]

Another method for comparison of results is the Kruskal-Wallis test~\cite{kruskal1952use,ostertagova2014methodology}. This test measures whether the median of several sets are~\textit{different}. We use this test to determine whether the pre-processing method had any effect on the performance of models for each disease. Since we are comparing 3 different pre-processing methods for 20 different heart conditions, we use the Holm-Bonferroni correction~\cite{holm1979simple,abdi2010holm}, and divide our initial alpha of 0.05 by 60 (since we run 3*20 tests), which gives us the new alpha $\alpha^*=8.33e^{-4}$. We reject the null hypothesis (i.e, the pre-processing method has an effect on the outcome) if $p<\alpha^*$. We've provided the raw p-value results in Table~\ref{tab:kruskal_wallis}, such that less conservative means of determining pre-processing significance can be applied.

\begin{table*}[!h]
\centering

  \caption{Kruskal-Wallis P-Value for importance of factor in diagnosis of each disease }
  
        \begin{tabular}{|l|r|r|r|}
        \hline
        Label / Heart Condition &             Sampling Rate & Bandpass & Normalizing\\\hline
1st degree av block                  &  3.50e-02  & 2.128e-01 & 2.74e-01  \\
abnormal QRS                         &  \textbf{3.56e-08}  & 7.514e-01 & \textbf{2.41e-08} \\
atrial fibrillation                  &  \textbf{6.53e-04}  & 9.828e-02 & 4.14e-03   \\
incomplete right bundle branch block &  \textbf{2.13e-11}  & 8.064e-01 & 3.30e-02   \\
left anterior fascicular block       &  2.06e-01  & 1.707e-01 & \textbf{2.00e-04}    \\
left axis deviation                  &  \textbf{2.33e-04}  & 3.918e-02 & \textbf{7.56e-06}   \\
left ventricular high voltage        &  2.84e-01  & 9.227e-02 & \textbf{1.44e-14}    \\
left ventricular hypertrophy         &  \textbf{6.34e-05}  & 6.495e-01  & \textbf{2.09e-12}   \\
myocardial infarction                &  \textbf{2.86e-10}  & 3.122e-02  & 1.62e-02  \\
myocardial ischemia                  &  \textbf{2.82e-04}  & 4.939e-01  & \textbf{2.80e-07}  \\
nonspecific st t abnormality         &  2.51e-02  & 5.428e-01 & 1.49e-02  \\
premature atrial contraction         &  \textbf{5.97e-08}  & 2.842e-01 & 3.28e-01  \\
right bundle branch block            &  1.89e-02  & 7.902e-01 & 2.38e-02  \\
sinus bradycardia                    &  \textbf{1.43e-05}  & 7.902e-01 & 9.28e-01  \\
sinus rhythm                         &  3.23e-01  & 9.566e-01 & 3.23e-01  \\
sinus tachycardia                    &  2.54e-03  & 2.685e-02 & 4.93e-01 \\
st depression                        &  6.40e-03  & 9.764e-01 & \textbf{4.43e-05}  \\
supraventricular tachycardia         &  9.73e-01  & 2.441e-01 & 5.42e-01 \\
t wave abnormal                      &  1.29e-01  & 7.262e-01 & 1.53e-01  \\
ventricular ectopics                 &  \textbf{4.02e-21}  & 2.357e-01 & 3.56e-01 \\\hline
        \end{tabular}
          \label{tab:kruskal_wallis}
\end{table*}

\begin{table*}[!htbp]
\centering

  \caption{P-value for Kruskal-Wallis test across conditions. Do sampling rate, band-passing, normalizing, and model architecture affect the F1 result?}
  \begin{tabular}{|r|r|r|r|}
  \hline
    \bfseries Sampling Rate & \bfseries Band-passing & \bfseries Normalization & \bfseries Model \\
  \hline
  3.65e-01 & 9.04e-01 & 2.00e-04 & 1.11e-4  \\
  \hline
  \end{tabular}
    \label{tab:kr_all_conditions}
\end{table*}
\section{Scaling/Down-sampling}
\label{sec:scaling}
The ECG signals in our dataset are recorded at 500Hz. We apply the  scaling rates of 0.1, 0.25, 0.5, 0.75, and 1, which down-samples the signals to sampling rates of 50Hz, 125Hz, 250Hz, 375Hz, and the unmodified 500Hz. After scaling, we applied ``min-max" norm, which we assumed to be the safest choice considering the popularity of normalization as a pre-processing method (see Section~\ref{sec:Norm}. The down-sampling method used here is Pytorch's ``nearest-exact" interpolation function~\cite{NEURIPS2019_9015}, which is a GPU implementation of the same algorithm by~\cite{van2014scikit} and the commonly used PIL libarary~\cite{clark2015pillow}. Effectively, each dataset has 5 different sampling rates, and each of the 3 models learns from these datasets by randomly selecting 80\% of the dataset for training, 10\% for validation, and 10\% for testing. In total, for each dataset, we run 15 experiments (3 architectures, and 5 different sampling rates) and for each experiment, we train and evaluate 20 times leading to 20 performance measures, where each performance measure is macro-average F1 score on the test set. We use the Wilcox-Holm post-hoc analysis described in the methodology (Section~\ref{sec:experiment}) to get the rank and relative performance for each model. 
\subsection{Down-sampling Results}
\label{sec:ds_results}
\begin{figure*}[htbp]
\centering

  \caption{Performance of the models for each disease at varying scaling rates. The 15 models are ranked against each other, higher MRR means better relative performance}
  {
  \resizebox{1\linewidth}{!}{\includegraphics{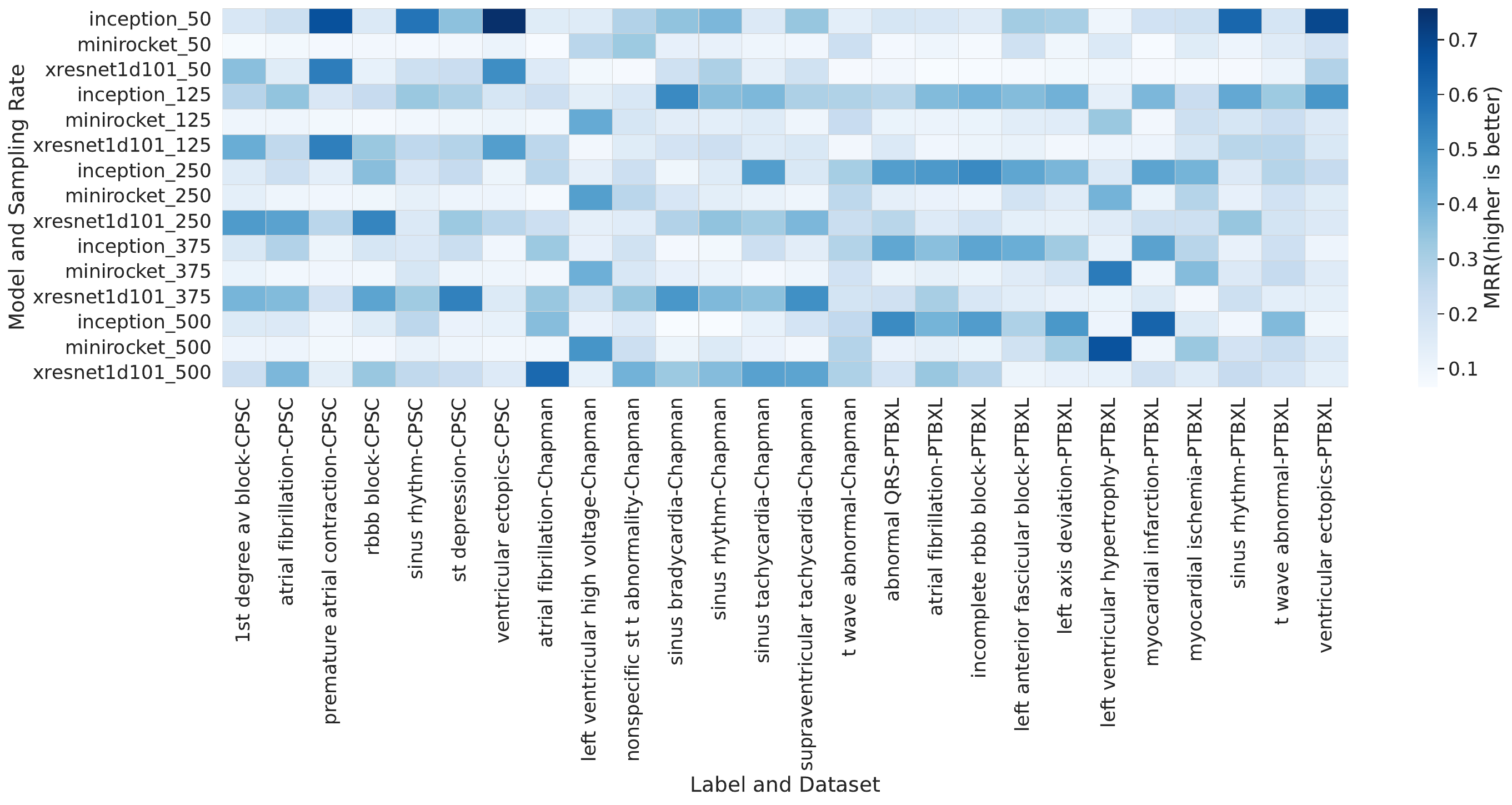}}
  }
    \label{fig:scaling_results_per_disease}
\end{figure*}

The MRR for each model when predicting each label is shown in Figure~\ref{fig:scaling_results_per_disease}. Here, no consistent pattern can be seen across diseases and scaling rates. However, some trends are observable: for example, Atrial fibrillation---a common label in the 3 datasets---deep learning models with higher sampling rates performed the best. Another interesting label is Sinus Rythm, also common in all 3 datasets,  where the inception network with lowest possible sampling rate clearly out-performs. Kruskal-Wallis p-values in Table~\ref{tab:kruskal_wallis} show that sampling rate is an important factor in 10 out of 20 diseases (the 10 significant p-values are highlighted in the Table under the ``Sampling Rate" column). From these observations we can conclude that sampling rate is an important factor for better diagnosis of \textit{some} heart conditions.

We also show the Spearman correlation between sampling rate and performance in Table~\ref{tab:corr}. There is a per-model breakdown of the correlation values, as well as the values when all datasets and all models are considered. Here, correlation measures the mapping between two continuous variables: sampling rates between 50-500Hz and F1 performance. Overall, when considering all models and all datasets, we see a negligible correlation of $0.008$. This suggests that sampling-rate is \textbf{not} an important factor in the overall results of our training. This weak relationship is also shown in   Table~\ref{tab:kr_all_conditions} where the Kruskal-Wallis test results across all diseases measured a p-value of 0.365
\begin{table}[htbp]
\centering

  \caption{Higher sampling rates required more training VRAM (in Gigabytes). Relative ratios should be considered, since absolute values are affected by window size, precision, batch size, etc.}
  \begin{tabular}{|r|r|r|r|}
  \hline
    \bfseries Hz & \bfseries inception & \bfseries xresent1d & \bfseries rocket\\
  \hline
    50  & 1.8       & 3.2     & 1.5    \\
    250 & 3.9       & 6.4     & 3.2    \\
    500 & 6.5       & 11.0    & 5.0   \\
  \hline
  \end{tabular}
  \label{tab:vram}
\end{table}

\textbf{Main Takeaway:} Based on these results, sampling rate does appear to be an important factor in diagnosis of multiple labels, however, it does not standout as a particularly important factor when training a multi-label model, unless some labels in the models are weighted differently than others.  This is a significant find as sampling-rate has a direct effect on the speed and hardware required for training deep-learning models. As shown in Table~\ref{tab:vram}, higher sampling rates have a higher VRAM cost, which affects training times and energy consumption. These results suggest that it is possible to reduce sampling rates from the standard 500Hz to values lower than 250Hz without major loss in over-all performance.

\begin{table}[htbp]
\centering

  \caption{Weak Spearman correlations are observed between F1 and sampling-rate (50-500Hz), and F1 and norm. ($0\xrightarrow{}$raw, $1\xrightarrow{}$min-max).}
  \begin{tabular}{|l|l|l|}
  \hline
   & \bfseries Rate & \bfseries Norm \\\hline
  
  Inception  & -5.23e-02 & -1.17e-01  \\
  MiniRocket &  2.70e-02 & -1.21e-02  \\
  xresnet    & 5.17e-02 & -9.50e-02 \\
  \hline
  All Models & 8.00e-03 & -7.41e-02\\
  \hline
  \end{tabular}
    \label{tab:corr}
\end{table}

\begin{figure*}[tbph]
\centering

  \caption{Performance of the models for each disease at varying HighPass frequencies. Higher MRR means better relative performance}
  {
  \resizebox{1\linewidth}{!}{\includegraphics{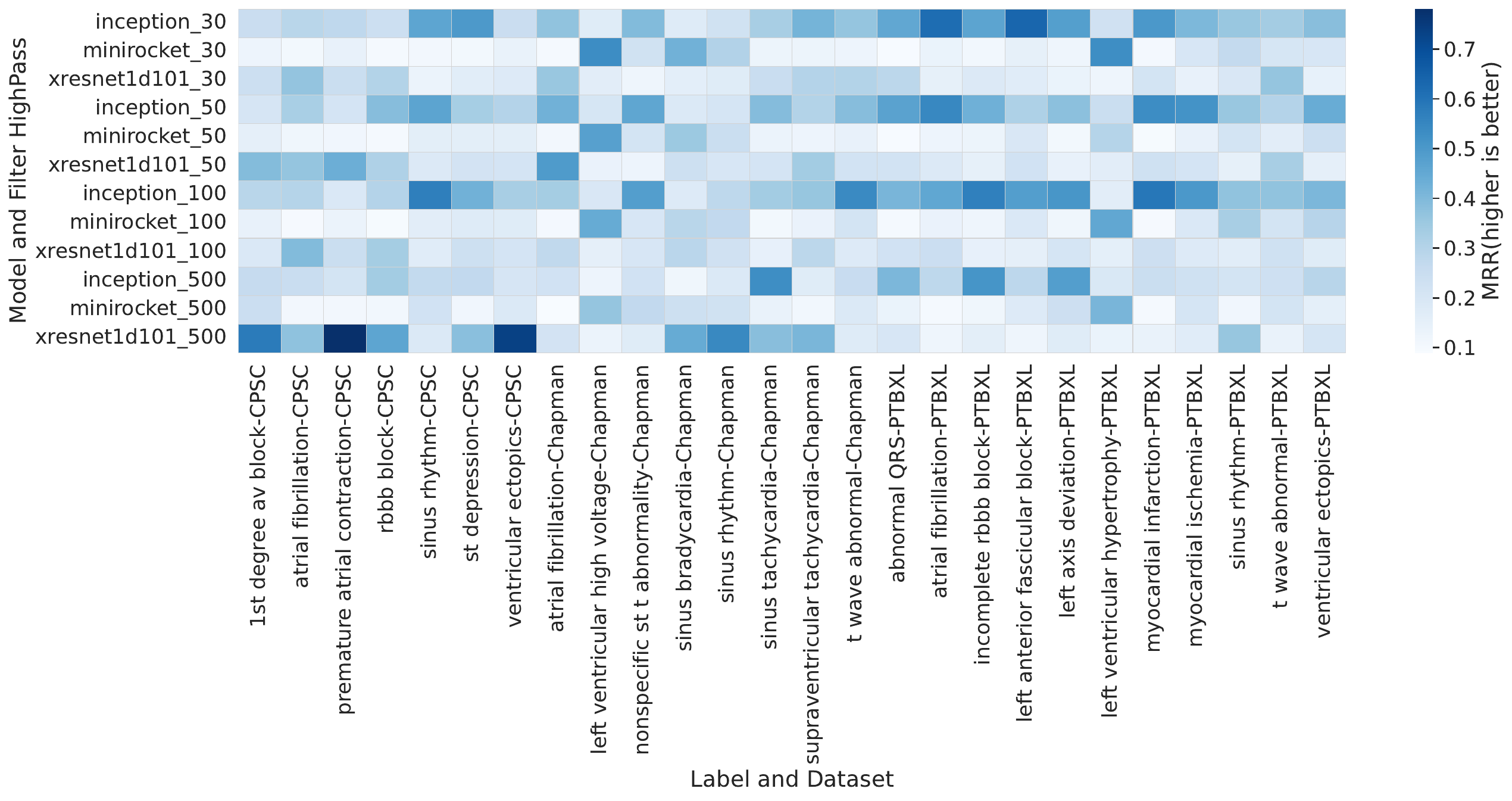}}
  }
    \label{fig:filtering_results_per_disease}
\end{figure*}

\begin{figure*}[tbph]
\centering

  \caption{Performance of the models for each disease using different normalization types. Higher MRR means better relative performance}
  {
  \resizebox{1\linewidth}{!}{\includegraphics{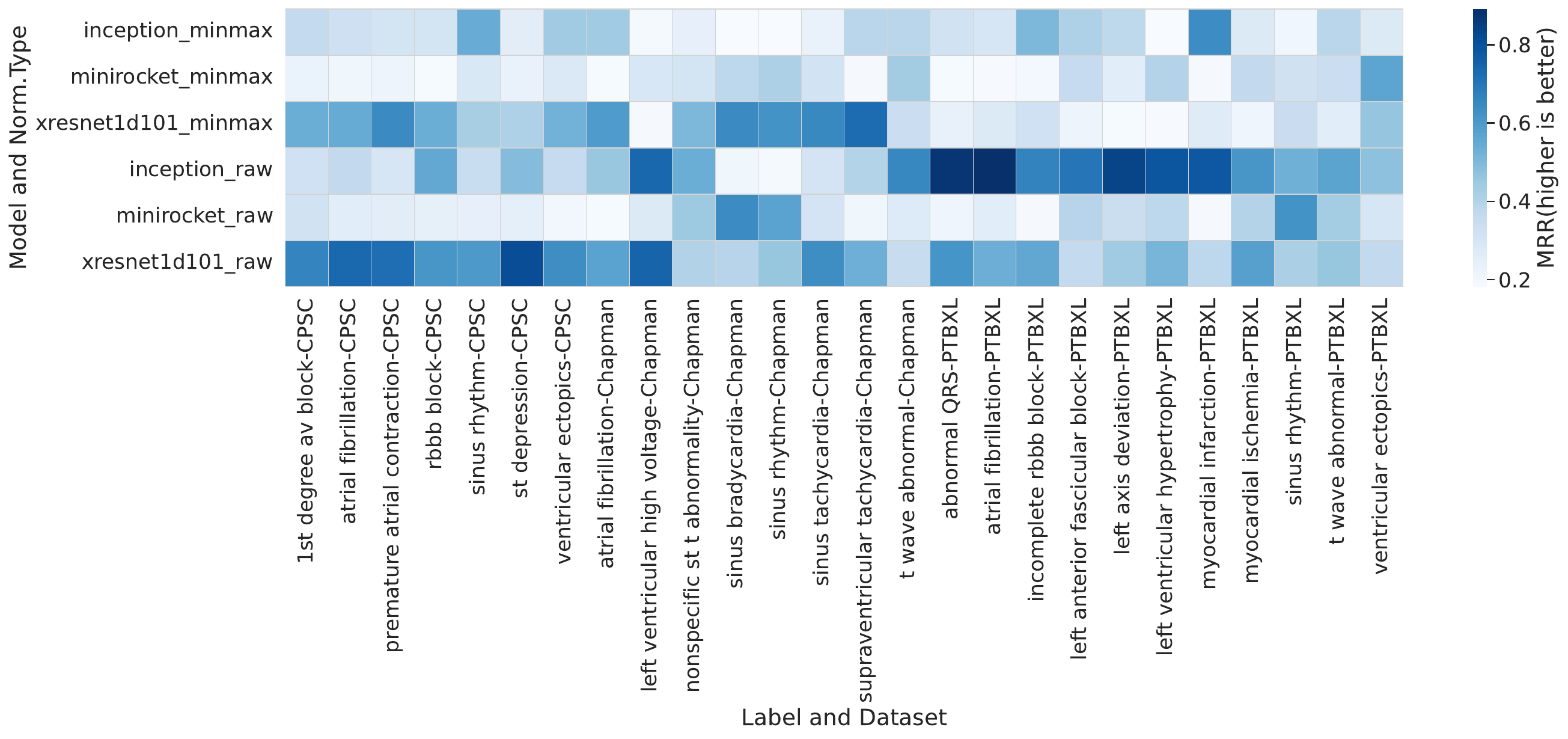}}
  }
    \label{fig:normalizing_results_per_disease}
\end{figure*}

\section{Band-Pass Filtering}
\label{sec:BandPass}
 A band-pass filter removes all frequencies outside its range, which can be an effective noise removal technique. The band-pass filter range applied to ECG data varies greatly in past works. Typically, the high-pass filter takes on a value of 0.05-1Hz, and the low-pass filter is applied somewhere between 30-150Hz~\cite{uwaechia2021comprehensive,luo2010review}. Is there an ``optimal" range for a band-pass filter in this context? Is band-pass filtering even necessary when using large datasets and large deep learning models?

In Section~\ref{sec:ds_results}, we discussed how down-sampling does not appear to have a major effect on the over-all model performance. Here, in order to speed up training, we downsample each dataset to 250Hz. We then apply 3 different band-pass filters to the data before training the 3 models. The 3 ranges are 1-30Hz, 1-50Hz, 1-100Hz. We compare these 3 pre-processing methods to see if the change in the low-pass filter frequency makes a difference in the final outcome, we then make a comparison to the case when no filtering was applied to the dataset. 

\subsection{Band-Pass Filtering Results}
The MRR for each model when predicting each label is shown in Figure~\ref{fig:filtering_results_per_disease}. The Kruskal-Wallis results in Table~\ref{tab:kruskal_wallis} do not indicate a difference between the performance of the 4 model groups in question (models trained using the 3 different band-passing functions, and models trained on the raw data). This is an interesting result which may explain the lack of consistency in the application of signal filtering in past ECG classification works. In addition, as shown in Table~\ref{tab:kr_all_conditions}, a p-value of 0.904 was measured by the Kruskal-Wallis test between F1 measures and different band-passing methods when considering all diseases. 
\\
\textbf{Main Takeaway:} When training on multiple labels from large, carefully curated datasets of ECGs, band-passing is not a necessary step. Band-pass filtering does not hurt the final outcomes, and can be applied at the GPU level with low overhead. Therefore in most situations, particularly when dealing with potentially noisy data (for example, where DC offset may not have been removed prior), it is advisable to use filtering. We cannot make any concrete statements about best cut-off frequencies, as it likely varies based on the recording hardware used and the cardio-vascular conditions being diagnosed. 
\section{Normalization}
\label{sec:Norm}
 What we refer to as "normalization" is the transformation of an input signal---which in this case is an ECG---to a distribution of values in the range of [0,1]~\cite{hong2022practical}. Here we use the ``Min-Max" normalization method, a commonly used pre-processing technique for ECG signals~\cite{uwaechia2021comprehensive,li2010robust,li2019identifying,fang2009human}. This technique uses the following formula where $Y[n]$ is the new amplitude for time-step $n$, $X[n]$ is the old amplitude, $min$ is the minimum value for $X[n]$, and $max$ is the maximum value for $X[n]$ \cite{uwaechia2021comprehensive}.

\[ Y[n] = \frac{x[n]-min}{max-min} \]

Here we ran the models an the datasets with no pre-processing, and compare the results to the models with only normalization applied.

\subsection{Normalization Results}
The MRR for each model when predicting each label is shown in Figure~\ref{fig:normalizing_results_per_disease}. P-values in Table~\ref{tab:kruskal_wallis} show normalization to be an important factor for 7 out of 20 diseases. For example, ``left ventricular high voltage" and ``left ventricular hypertrophy" performed better without normalization in all 3 models. To calculate the correlation values between F1 and the normalization type in Table~\ref{tab:corr}, we mapped the label of raw datasets to 0, and min-maxed datasets to 1. The correlation values have a slight negative bias, indicating that when normalization was important, more often than not models training on raw ECGs performed better. In addition, as shown in Table~\ref{tab:kr_all_conditions}, a small p-value of 0.0002 was measured by the Kruskal-Wallis test between F1 measures and the two normalization groups; this indicates that normalization (or lack thereof) was overall an important factor.

\textbf{Main Takeaway} Surprisingly, Min-Max normalization has a slight negative impact on the performance relative to raw ECGs. We cannot discourage the use of all normalization methods as it is only one of the common normalization methods used \cite{uwaechia2021comprehensive}. However, these results indicate that staple pre-processing methods in other fields of deep-learning cannot be presumed effective for ECG classification.
\section{Discussion}
Prior to this work, we found no consistency in the pre-processing methods used in classification of cardiovascular conditions using ECG. To address this gap, we trained 3 different classifiers on 3 datasets of ECGs and examined the effect of 3 pre-processing methods on the performance of the models in classifying various heart conditions. Our findings suggest that the performance of different pre-processing and architecture combinations varies depending on the condition of interest. This would explain the lack of consistency in previous works, where datasets, label weights, and performance measures vary. 

We found rates as low as 50Hz can yield comparable if not better results than the commonly used 500Hz sampling rate. This is significant as smaller sampling rates can drastically reduce time and hardware requirements for training models. We also found min-max normalization to be slightly detrimental, and band-passing to make no measurable difference in model performance. Beyond these findings, the overall takeaway of this work is that there are no ``one-size-fits-all'' solutions to the problem of ECG pre-processing, and the ``optimal'' best pre-processing method is label and architecture dependent. While this result may seem  unsatisfactory, at the very least this work shows that researchers aiming to utilize SOTA, high-performing models for different labels and datasets cannot assume that the pre-processing method used by other works will be optimal. This work also highlights the need for an effective, automated pre-processing step in the ECG classification domain (or time-series in general), which appears to be an open problem.

Here, we ignored the possible interdependence between pre-processing methods. An approach which considers the \textit{sequence} in which the methods are applied would exponentially increase the---already costly---time and hardware requirements for this work. Additionally, past works rarely discuss the sequencing in which their pre-processing methods were applied, making their replication of all the more difficult. 

Many recent works have noted the higher accuracy of an ensemble model with various architectures relative to any single model~\cite{uwaechia2021comprehensive,hong2022practical,IsmailFawaz2018deep,strodthoff2020deep,chen2020detection}, although the diminishing returns of accuracy vs training time and hardware costs should also be considered. Our findings here suggest that the use of various architectures \textit{as well} as different pre-processing methods could yield slight improvements in benchmarks.  However, taking such an approach would be brute-forcing the problem, rather than solving it.

\bibliographystyle{IEEEbib}
\bibliography{strings,refs}

\begin{thebibliography}{10}

\bibitem{maleki2014use}
Neda~Dianati Maleki, Arash~Ehteshami Afshar, and Paul~W Armstrong,
\newblock ``Use of electrocardiogram indices of myocardial ischemia for risk
  stratification and decision making of reperfusion strategies,''
\newblock {\em Journal of Electrocardiology}, vol. 47, no. 4, pp. 520--524,
  2014.

\bibitem{uwaechia2021comprehensive}
Anthony~Ngozichukwuka Uwaechia and Dzati~Athiar Ramli,
\newblock ``A comprehensive survey on ecg signals as new biometric modality for
  human authentication: Recent advances and future challenges,''
\newblock {\em IEEE Access}, 2021.

\bibitem{reyna2021will}
Matthew~A Reyna, Nadi Sadr, Erick A~Perez Alday, Annie Gu, Amit~J Shah, Chad
  Robichaux, Ali~Bahrami Rad, Andoni Elola, Salman Seyedi, Sardar Ansari,
  et~al.,
\newblock ``Will two do? varying dimensions in electrocardiography: the
  physionet/computing in cardiology challenge 2021,''
\newblock in {\em 2021 Computing in Cardiology (CinC)}. IEEE, 2021, vol.~48,
  pp. 1--4.

\bibitem{reyna4issues}
Matthew~A Reyna, Nadi Sadr, Erick A~Perez Alday, Annie Gu, Amit~J Shah, Chad
  Robichaux, Ali~Bahrami Rad, Andoni Elola, Salman Seyedi, Sardar Ansari,
  et~al.,
\newblock ``Issues in the automated classification of multilead ecgs using
  heterogeneous labels and populations,''
\newblock {\em personnel}, vol. 4, pp. 5, 2021.

\bibitem{ribeiro2020automatic}
Ant{\^o}nio~H Ribeiro, Manoel~Horta Ribeiro, Gabriela~MM Paix{\~a}o, Derick~M
  Oliveira, Paulo~R Gomes, J{\'e}ssica~A Canazart, Milton~PS Ferreira, Carl~R
  Andersson, Peter~W Macfarlane, Wagner Meira~Jr, et~al.,
\newblock ``Automatic diagnosis of the 12-lead ecg using a deep neural
  network,''
\newblock {\em Nature communications}, vol. 11, no. 1, pp. 1--9, 2020.

\bibitem{chen2020detection}
Tsai-Min Chen, Chih-Han Huang, Edward~SC Shih, Yu-Feng Hu, and Ming-Jing Hwang,
\newblock ``Detection and classification of cardiac arrhythmias by a
  challenge-best deep learning neural network model,''
\newblock {\em Iscience}, vol. 23, no. 3, pp. 100886, 2020.

\bibitem{zhao2020adaptive}
Zhibin Zhao, Hui Fang, Samuel~D Relton, Ruqiang Yan, Yuhong Liu, Zhijing Li,
  Jing Qin, and David~C Wong,
\newblock ``Adaptive lead weighted resnet trained with different duration
  signals for classifying 12-lead ecgs,''
\newblock in {\em 2020 Computing in Cardiology}. IEEE, 2020, pp. 1--4.

\bibitem{hong2022practical}
Shenda Hong, Wenrui Zhang, Chenxi Sun, Yuxi Zhou, and Hongyan Li,
\newblock ``Practical lessons on 12-lead ecg classification: Meta-analysis of
  methods from physionet/computing in cardiology challenge 2020,''
\newblock {\em Frontiers in Physiology}, p. 2505, 2022.

\bibitem{natarajan2020wide}
Annamalai Natarajan, Yale Chang, Sara Mariani, Asif Rahman, Gregory Boverman,
  Shruti Vij, and Jonathan Rubin,
\newblock ``A wide and deep transformer neural network for 12-lead ecg
  classification,''
\newblock in {\em 2020 Computing in Cardiology}. IEEE, 2020, pp. 1--4.

\bibitem{zhu2020classification}
Zhaowei Zhu, Han Wang, Tingting Zhao, Yangming Guo, Zhuoyang Xu, Zhuo Liu, Siqi
  Liu, Xiang Lan, Xingzhi Sun, and Mengling Feng,
\newblock ``Classification of cardiac abnormalities from ecg signals using
  se-resnet,''
\newblock in {\em 2020 Computing in Cardiology}. IEEE, 2020, pp. 1--4.

\bibitem{oppelt2020combining}
Maximilian~P Oppelt, Maximilian Riehl, Felix~P Kemeth, and Jan Steffan,
\newblock ``Combining scatter transform and deep neural networks for multilabel
  electrocardiogram signal classification,''
\newblock in {\em 2020 Computing in Cardiology}. IEEE, 2020, pp. 1--4.

\bibitem{hasani2020classification}
Hosein Hasani, Adeleh Bitarafan, and Mahdieh~Soleymani Baghshah,
\newblock ``Classification of 12-lead ecg signals with adversarial multi-source
  domain generalization,''
\newblock in {\em 2020 Computing in Cardiology}. IEEE, 2020, pp. 1--4.

\bibitem{goldberger2000physiobank}
Ary~L Goldberger, Luis~AN Amaral, Leon Glass, Jeffrey~M Hausdorff, Plamen~Ch
  Ivanov, Roger~G Mark, Joseph~E Mietus, George~B Moody, Chung-Kang Peng, and
  H~Eugene Stanley,
\newblock ``Physiobank, physiotoolkit, and physionet: components of a new
  research resource for complex physiologic signals,''
\newblock {\em circulation}, vol. 101, no. 23, pp. e215--e220, 2000.

\bibitem{alday2020classification}
Erick A~Perez Alday, Annie Gu, Amit~J Shah, Chad Robichaux, An-Kwok~Ian Wong,
  Chengyu Liu, Feifei Liu, Ali~Bahrami Rad, Andoni Elola, Salman Seyedi,
  et~al.,
\newblock ``Classification of 12-lead ecgs: the physionet/computing in
  cardiology challenge 2020,''
\newblock {\em Physiological measurement}, vol. 41, no. 12, pp. 124003, 2020.

\bibitem{luo2010review}
Shen Luo and Paul Johnston,
\newblock ``A review of electrocardiogram filtering,''
\newblock {\em Journal of electrocardiology}, vol. 43, no. 6, pp. 486--496,
  2010.

\bibitem{rukundo2012nearest}
Olivier Rukundo and Hanqiang Cao,
\newblock ``Nearest neighbor value interpolation,''
\newblock {\em arXiv preprint arXiv:1211.1768}, 2012.

\bibitem{Makowski2021neurokit}
Dominique Makowski, Tam Pham, Zen~J. Lau, Jan~C. Brammer, Fran{\c{c}}ois
  Lespinasse, Hung Pham, Christopher Schölzel, and S.~H.~Annabel Chen,
\newblock ``{NeuroKit}2: A python toolbox for neurophysiological signal
  processing,''
\newblock {\em Behavior Research Methods}, vol. 53, no. 4, pp. 1689--1696, feb
  2021.

\bibitem{lyons1997understanding}
Richard~G Lyons,
\newblock {\em Understanding digital signal processing, 3/E},
\newblock Pearson Education India, 1997.

\bibitem{bhanja2018impact}
Samit Bhanja and Abhishek Das,
\newblock ``Impact of data normalization on deep neural network for time series
  forecasting,''
\newblock {\em arXiv preprint arXiv:1812.05519}, 2018.

\bibitem{tsai}
Ignacio Oguiza,
\newblock ``tsai - a state-of-the-art deep learning library for time series and
  sequential data,'' Github, 2022.

\bibitem{szegedy2017inception}
Christian Szegedy, Sergey Ioffe, Vincent Vanhoucke, and Alexander~A Alemi,
\newblock ``Inception-v4, inception-resnet and the impact of residual
  connections on learning,''
\newblock in {\em Thirty-first AAAI conference on artificial intelligence},
  2017.

\bibitem{ismail2020inceptiontime}
Hassan Ismail~Fawaz, Benjamin Lucas, Germain Forestier, Charlotte Pelletier,
  Daniel~F Schmidt, Jonathan Weber, Geoffrey~I Webb, Lhassane Idoumghar,
  Pierre-Alain Muller, and Fran{\c{c}}ois Petitjean,
\newblock ``Inceptiontime: Finding alexnet for time series classification,''
\newblock {\em Data Mining and Knowledge Discovery}, vol. 34, no. 6, pp.
  1936--1962, 2020.

\bibitem{dempster2021minirocket}
Angus Dempster, Daniel~F Schmidt, and Geoffrey~I Webb,
\newblock ``Minirocket: A very fast (almost) deterministic transform for time
  series classification,''
\newblock in {\em Proceedings of the 27th ACM SIGKDD Conference on Knowledge
  Discovery \& Data Mining}, 2021, pp. 248--257.

\bibitem{ruiz2021great}
Alejandro~Pasos Ruiz, Michael Flynn, James Large, Matthew Middlehurst, and
  Anthony Bagnall,
\newblock ``The great multivariate time series classification bake off: a
  review and experimental evaluation of recent algorithmic advances,''
\newblock {\em Data Mining and Knowledge Discovery}, vol. 35, no. 2, pp.
  401--449, 2021.

\bibitem{Strodthoff2021}
Nils Strodthoff, Patrick Wagner, Tobias Schaeffter, and Wojciech Samek,
\newblock ``Deep learning for ecg analysis: Benchmarks and insights from
  ptb-xl,''
\newblock {\em IEEE Journal of Biomedical and Health Informatics}, vol. 25, no.
  5, pp. 1519--1528, 2021.

\bibitem{strodthoff2020deep}
Nils Strodthoff, Patrick Wagner, Tobias Schaeffter, and Wojciech Samek,
\newblock ``Deep learning for ecg analysis: Benchmarks and insights from
  ptb-xl,''
\newblock {\em IEEE Journal of Biomedical and Health Informatics}, vol. 25, no.
  5, pp. 1519--1528, 2020.

\bibitem{liu2018open}
Feifei Liu, Chengyu Liu, Lina Zhao, Xiangyu Zhang, Xiaoling Wu, Xiaoyan Xu,
  Yulin Liu, Caiyun Ma, Shoushui Wei, Zhiqiang He, et~al.,
\newblock ``An open access database for evaluating the algorithms of
  electrocardiogram rhythm and morphology abnormality detection,''
\newblock {\em Journal of Medical Imaging and Health Informatics}, vol. 8, no.
  7, pp. 1368--1373, 2018.

\bibitem{zheng202012}
Jianwei Zheng, Jianming Zhang, Sidy Danioko, Hai Yao, Hangyuan Guo, and Cyril
  Rakovski,
\newblock ``A 12-lead electrocardiogram database for arrhythmia research
  covering more than 10,000 patients,''
\newblock {\em Scientific Data}, vol. 7, no. 1, pp. 1--8, 2020.

\bibitem{wagner2020ptb}
Patrick Wagner, Nils Strodthoff, Ralf-Dieter Bousseljot, Dieter Kreiseler,
  Fatima~I Lunze, Wojciech Samek, and Tobias Schaeffter,
\newblock ``Ptb-xl, a large publicly available electrocardiography dataset,''
\newblock {\em Scientific data}, vol. 7, no. 1, pp. 1--15, 2020.

\bibitem{IsmailFawaz2018deep}
Hassan Ismail~Fawaz, Germain Forestier, Jonathan Weber, Lhassane Idoumghar, and
  Pierre-Alain Muller,
\newblock ``Deep learning for time series classification: a review,''
\newblock {\em Data Mining and Knowledge Discovery}, vol. 33, no. 4, pp.
  917--963, 2019.

\bibitem{kruskal1952use}
William~H Kruskal and W~Allen Wallis,
\newblock ``Use of ranks in one-criterion variance analysis,''
\newblock {\em Journal of the American statistical Association}, vol. 47, no.
  260, pp. 583--621, 1952.

\bibitem{ostertagova2014methodology}
Eva Ostertagova, Oskar Ostertag, and Jozef Kov{\'a}{\v{c}},
\newblock ``Methodology and application of the kruskal-wallis test,''
\newblock in {\em Applied Mechanics and Materials}. Trans Tech Publ, 2014, vol.
  611, pp. 115--120.

\bibitem{holm1979simple}
Sture Holm,
\newblock ``A simple sequentially rejective multiple test procedure,''
\newblock {\em Scandinavian journal of statistics}, pp. 65--70, 1979.

\bibitem{abdi2010holm}
Herv{\'e} Abdi,
\newblock ``Holm’s sequential bonferroni procedure,''
\newblock {\em Encyclopedia of research design}, vol. 1, no. 8, pp. 1--8, 2010.

\bibitem{NEURIPS2019_9015}
Adam Paszke, Sam Gross, Francisco Massa, Adam Lerer, James Bradbury, Gregory
  Chanan, Trevor Killeen, Zeming Lin, Natalia Gimelshein, Luca Antiga, Alban
  Desmaison, Andreas Kopf, Edward Yang, Zachary DeVito, Martin Raison, Alykhan
  Tejani, Sasank Chilamkurthy, Benoit Steiner, Lu~Fang, Junjie Bai, and Soumith
  Chintala,
\newblock ``Pytorch: An imperative style, high-performance deep learning
  library,''
\newblock in {\em Advances in Neural Information Processing Systems 32},
  H.~Wallach, H.~Larochelle, A.~Beygelzimer, F.~d\textquotesingle
  Alch\'{e}-Buc, E.~Fox, and R.~Garnett, Eds., pp. 8024--8035. Curran
  Associates, Inc., 2019.

\bibitem{van2014scikit}
Stefan Van~der Walt, Johannes~L Sch{\"o}nberger, Juan Nunez-Iglesias,
  Fran{\c{c}}ois Boulogne, Joshua~D Warner, Neil Yager, Emmanuelle Gouillart,
  and Tony Yu,
\newblock ``scikit-image: image processing in python,''
\newblock {\em PeerJ}, vol. 2, pp. e453, 2014.

\bibitem{clark2015pillow}
Alex Clark,
\newblock ``Pillow (pil fork) documentation,'' 2015.

\bibitem{li2010robust}
Ming Li and Shrikanth Narayanan,
\newblock ``Robust ecg biometrics by fusing temporal and cepstral
  information,''
\newblock in {\em 2010 20th International Conference on Pattern Recognition}.
  IEEE, 2010, pp. 1326--1329.

\bibitem{li2019identifying}
Yaoguang Li and Wei Cui,
\newblock ``Identifying the mislabeled training samples of ecg signals using
  machine learning,''
\newblock {\em Biomedical signal processing and control}, vol. 47, pp.
  168--176, 2019.

\bibitem{fang2009human}
Shih-Chin Fang and Hsiao-Lung Chan,
\newblock ``Human identification by quantifying similarity and dissimilarity in
  electrocardiogram phase space,''
\newblock {\em Pattern Recognition}, vol. 42, no. 9, pp. 1824--1831, 2009.

\end{thebibliography}

\end{document}